\documentclass[aps,pre,twocolumn,showpacs]{revtex4}
\usepackage{graphicx} \usepackage{amssymb} \usepackage{amsmath}
\begin{document}

\title{Damage Growth in Random Fuse Networks}
\author{F. Reurings and M. J. Alava}
\affiliation{Helsinki University of Technology, Laboratory of Physics,
P.O.Box 1100, FIN-02015 HUT, Finland}

\date{\today}

\begin{abstract}
The correlations among elements that break in random fuse 
network fracture are studied, for disorder
strong enough to allow for volume damage before final
failure. The growth of microfractures is found to be uncorrelated
above a lengthscale, that increases as the the final breakdown
is approached. Since the fuse network strength decreases
with sample size, asymptotically the process resembles
more and more mean-field-like (``democratic fiber bundle'') fracture.
This is found from the microscopic dynamics of avalanches or
microfractures, from a study of damage localization via entropy,
and from the final damage profile. In particular, the last one
is statistically constant, except exactly at the final crack
zone (in contrast to recent results by Hansen et al., Phys.
Rev. Lett.  \textbf{90}, 045504 (2003)), in spite of the
fact that the fracture surfaces are self-affine.
\end{abstract}

\maketitle
	
\section{Introduction}\label{intro}
The scaling properties of fracture processes continue
to attract interest from the statistical mechanics
community. Key quantities are the geometric properties
of fracture surfaces and statistics of acoustic emission,
or, in analogy to other systems,  ``crackling noise''.
The point is that in failure of brittle materials the
elastic energy of a sample is released in bursts. These
``avalanches'' often turn out to have scale-invariant
statistics with respect to e.g. the probability distribution
of the released energy \cite{Lockner,Petri2,Ciliberto,penn,sal1}. 
Likewise, crack surfaces are often
self-affine (with an empirical roughness exponent $\zeta$) 
\cite{Mandel,Bourev,bouprl}.
The understanding of the origins of such critical-like
statistics would perhaps be of interest to engineers
(``how to make tougher materials'') but would also
mean the solution of a very complicated many-particle system.

In this respect, among the simplest models that are available
are mean-field like fiber bundle models (FBM) \cite{hansen,Zapperi}
and random fuse networks (RFN's) \cite{review,Dux}.
The former describe democratic or global load sharing, and thus
do not have anything close to the stress enhancements of real
cracks (though one can introduce local load sharing to fiber bundles, 
and interpolate between these two limits as well). Such stress
effects are to be found in a natural
way in fuse networks, that simplify real elasticity by considering
the electrostatic analogy. RFN's have two natural limits: weak
disorder, when cracks are nucleated quickly and brittle failure
takes place without much precursor activity, and strong disorder
(without infinitely strong elements), where {\em damage}
develops before macroscopic failure \cite{review,Kahng}.

The same signatures are found in the latter, RFN, case, that also
characterize experimental systems: rough, self-affine cracks
and microcracking that corresponds to the acoustic emission.
The roughness exponent is in the proximity of $\zeta_{RFN} \sim 0.7$,
in 2d, tantalizingly close to the minimum energy surface exponent, exactly
2/3. This result holds also for e.g. 'weak' disorder 
\cite{Hansen,rai,sep} and is close to what is seen in experiments 
\cite{Kertesz,Maloy,Rosti,sal2}. The damage
develops in avalanches \cite{Zapperi,sal1,Zapperi2}, 
in analogy to democratic FBM's \cite{pitka},
so that one has for the probability distribution of the number of
fuses  ($\Delta)$ blown in one 'event', for current-control, 
\begin{equation}
\label{eq:DeltaP}
P(\Delta)\sim\Delta^{-5/2}.
\end{equation}
The corresponding AE energy exponent is about $\beta =1.7$ 
\cite{sal1,minozzi}, as expected
based on the exponent relation $\beta = (5/2+1)/2$ \cite{minozzi}. 
However, even
for strong disorder finally stress enhancements come into play,
and the sample fails catastrophically, with the elastic modulus
(conductivity, in the RFN case) having a first-order drop.

\begin{figure}
\begin{center}
\includegraphics[scale=0.5]{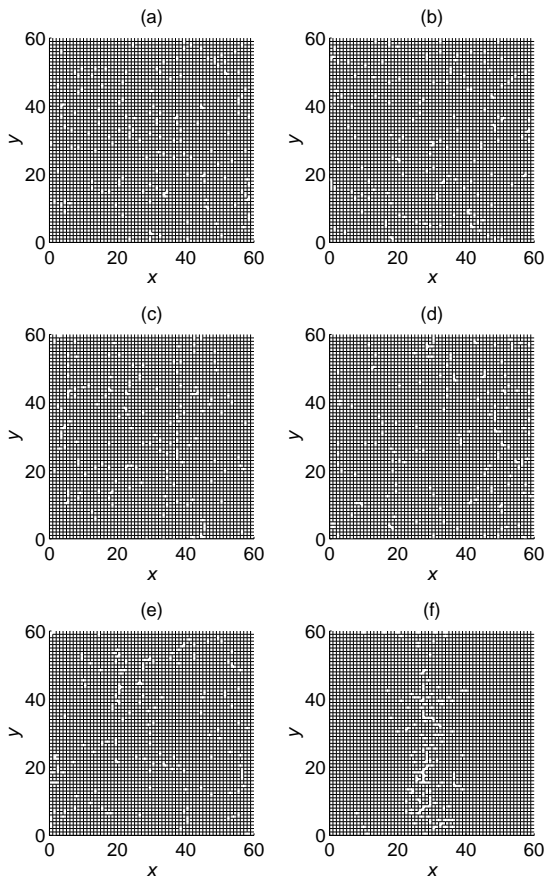}
\caption{Snapshots of subsequent damage patters in a failure
of a RFN (each sub-plot having the same number of failed fuses,
separately). $R=1$, ``strong disorder'', $L=60$.}
\label{fg:hoocee}
\end{center}
\end{figure}

The purpose of this article is to investigate the development
of the damage, between the MF-limit valid at the very initial
stages of fracture and the final critical crack growth. Figure
\ref{fg:hoocee} shows an example of the transition. The subplots
depict the individual fuses that fail in subsequent parts of
a stress-strain- (or current-voltage) history. Clearly, initially
the damage is random (unless proven differently by more sophisticated
analysis), and in the last panel it concentrates on the vicinity
and at the final crack.

In this respect, it is an important question how the pre-critical
damage reflects the self-affine properties of the final fracture
surface. Recently, Hansen and co-workers have attempted to relate
its formation to a self-consistently developed damage profile that
extends over all the sample \cite{Hansen2,ramstad}.
The scaling of the profile with the system size would then explain
the roughness and its exponent.
Clearly, this should also be visible in the dynamics of failure also prior
to the end of the process. Another analogy is given by dynamics
in dipolar
random field magnets, which can account for the symmetry breaking
(as signaled by the formation of the final crack) due to 
shielding in the direction of the external voltage and for 
stress enhancements that drive cracks mostly perpendicular
to it \cite{dipolar}.

We study these aspects by concentrating on two kinds
of quantities: those that characterize the spatial distribution
of damage in samples, and those that analyze the temporal
correlations in individual failure events (as e.g. during
an avalanche, or series of fuse failures due to the increase
of a control parameter). Section II considers the former,
and uses as the main tool entropy, comparing the damage
integrated over windows of time and/or space to that
in, spatially, completely random damage formation. We also
study the damage profiles of completely failed samples.
From both kinds of analysis emerges a picture of crack
development, that is mean-field -like beyond a finite
interaction range, and until the final breakdown is
induced by rare event statistics \cite{Dux}. This in particular
includes the fact that except in the ``fracture process
zone'', i.e. in the vicinity of the final crack, the
damage is statistically homogeneous. Thus in this
particular case of RFN's the theory proposed by Hansen
et al. seems unlikely to be the explanation for the
self-affine geometry of cracks.

In Section III, the internal dynamics of avalanches is 
considered. We look at the probability distributions
and average values of ``jumps'' (relative changes in the
position of subsequent failures). It transpires
that there is a smooth development, in which the
these quantities exhibit a cross-over from the
FBM/MF-like lack of spatial inhomogeneity towards
localized crack growth within a scale $\xi$. 
This resembles some observations
by Curtin about critical damage clusters in a more
elaborate fiber bundle-type model: the dynamics is
based on democratic load sharing in spite of the
presence of stress enhancements \cite{Curtin}. It also pertains
to the question of the existence of ``representative
volume elements'' \cite{van} or coarse-graining in fuse networks 
\cite{deplace}, related to the general question of how to
account for microscopic dynamics and phenomena with coarse-grained
variables and equations.
Beyond any such correlation length $\xi$ as may exist 
within avalanches the network looks homogeneous, if in
addition the damage density is statistically homogeneous.
Finally, we finish the paper in Sec. IV with a summary and 
some open prospects.

\section{Distribution of damage}

The RFN's, as electrical analogues of (quasi-)brittle
media consist of fuses with a linear voltage-current relationship 
until a breakdown current $i_b$. A stress-strain test can
be done by using adiabatic fracture iterations: the current
balance is solved, and at each round the most strained fuse is
chosen according to $\min(i_j / j_{c,j})$, where $i_j$ is the local 
current and $j_{c,j}$ the local threshold). Currents and voltages
are solved by the conjugate-gradient method. In the following
we use for $P(i_c )$ a flat distribution $P(i_c)= [1-R,1+R]$,
with the disorder parameter $R$ chosen as unity. The simulations
are done in 2d, in the (10) lattice orientation, with periodic
boundary conditions in the transverse direction (y). Square systems
upto $100^2$ have been studied; notice that the damage is in
practice volume-like, and thus thousands of iterations are needed
per a single system for $L \sim 100$.

Studies of the break-down current $I_b$ as a function affirm
the expected outcome of a logarithmic scaling \cite{Dux,sep}, resulting
from extremal statistics ($I_b \sim L/\ln{L}$). This implies in
the mean-field limit that $n_b\sim\frac{L^2}{\log L}$, where
$n_b$ is the average number of broken fuses in a system.

To analyze the spatial of distribution of damage it is useful
first to take note of the fact that in the latter stages
the system behaves anisotropically: just before the formation
of a critical crack the spatial density of the broken fuses
should be a stochastic variable, with a constant mean in the
transverse direction to the external voltage. However, along
the voltage direction differences may ensue. To study such
trends in the damage mechanics and the localization 
we consider the entropy of the damage averaged over y in
each sample (this is in analogy to the procedure used with
AE experiments of Guarino
et al. \cite{Ciliberto}). The network is divided along the current flow
direction into sections, and the entropy $S$  defined as
\begin{equation}
\label{entropydef}
S=-\sum_i q_i\ln q_i,
\end{equation}
where $q_i$ is the fraction of burned fuses in section $i$. 
$S$ is normalized by $S_e$, the entropy of a random, on the
average homogeneous
distribution of failures (of equal total damage). 
Thus the extreme limits are zero and
unity, corresponding to completely localized damage and complete
random one, respectively. The final crack extends between
$y=0$ and $y=L$, and a sensible choice is to use for the section
width $\delta x$ a value larger than the typical interface width $w$,
\begin{equation}
\label{eq:wdef}
w=\langle(h_y-\bar{h})^2\rangle^{1/2},
\end{equation}
where $h_y$ denotes the crack location and $\bar{h}$ its mean position
in the $x$-direction.
Since $w\sim L^\zeta$, with $\zeta<1$, it is clear that using
a constant number of sections will with increasing $L$ localize
the fracture zone either entirely inside one, or between two neighboring
ones. For better statistics it is preferable to have $\delta x >>1$,
though the interpretation is perhaps more difficult than for the extreme
value $\delta x =1$, say.
The width of the sections used in computing the entropies was chosen to be 
$\delta x=L/10$.

In this discrete form, the entropy reads
\begin{equation}
S=-\sum_{i=1}^k\frac{n_i}{N}\ln\frac{n_i}{N},
\end{equation}
where $k=L/\delta x$ is the number of sections, $n_i$ the number of
burnt fuses in the $i$'th section, and $N=\sum_{i=1}^kn_i$ is 
again the total. Note that the absolute value of $S$ is dependent
on the choice for $\delta x$. $S$ can now be used to consider
different parts of the stress-strain curve, separately, or the
final damage pattern.

\begin{figure}
\begin{center}
\includegraphics[scale=0.35]{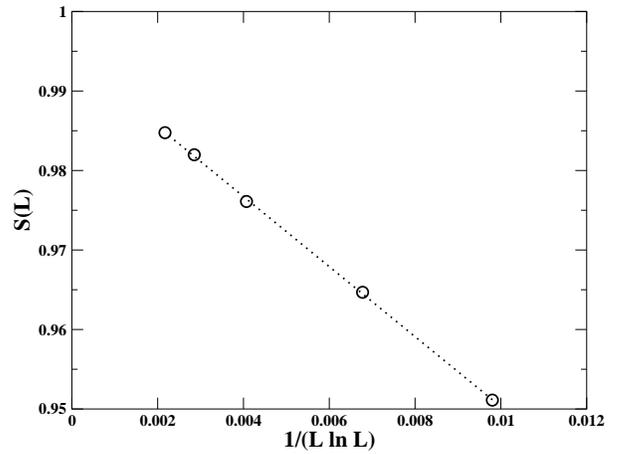}
\caption{\small $S$ vs. $1/(L\log L)$ in networks with strong disorder ($R=1$). The straight line is a linear least squares fit that intersects the $S$-axis at $S=0.9943$.}
\label{fg:SvsL}
\end{center}
\end{figure}

Fig. \ref{fg:SvsL} shows the total entropy versus system size.
The best kind of linearity with regards the data
is obtained with a scaling variable 
$1/(L\log L)$. This can be considered with the following Ansatz. Assume, that
the fractures are distributed otherwise randomly ($n_{b,i}$), except the 
one containing the final crack, which has $n_{b,k}+\Delta n_{b,k}$.
Take  $\Delta n_b \ll \langle n_b \rangle$, which implies 
approximately $S\approx1-\ln k(\Delta n_{b,k}/n_b)$. 
Noticing the logarithmic
scaling of damage, it follows that $S\sim-1/(L\log L)$,
if and only if $\Delta n_{b,k}$ scales as $\Delta n_{b,k}\sim L/(\log L)^2$. 
We have not checked explicitly that this holds; note that the
fracture surface, being self-affine, is then supposed to contain
$L^a$ fuses, with $1<a\ll2$, but the fracture process zone
itself, contains other damage (broken fuses) contributing
to $\Delta n_b$ (see Fig.~1 again, and the last panel in particular). 
In any case, it is obvious that the entropy
{\em increases} with system size, indicating more and more
completely random damage. The limiting value of $S$ is slightly below
unity; it is hard to say whether this difference is due to
the choice used for computing $S$ or a real one. 

\begin{figure}
\begin{center}
\includegraphics[scale=0.35]{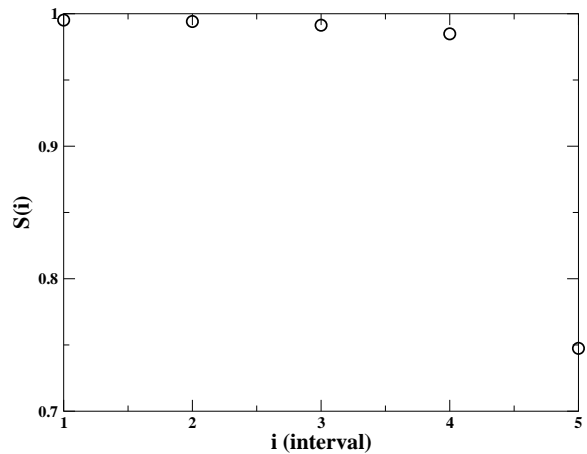}
\caption{\small Entropy $S$ versus time interval $\Delta n_i$. $L=100$, $R=1$. Average over 20 realizations.}
\label{fg:Svsdn}
\end{center}
\end{figure}

In figure \ref{fg:Svsdn} an example is shown of how the
damage actually localizes when the fracture process is
divided into sequential slices. It is clear that initially
most of the fuses break randomly, and only in the last one
strong localization takes place. Similar sampling can be
done also with the external voltage or current as control
parameters, the difference between these two being that
the fracture is more abrupt in the latter case.

\begin{figure}
\begin{center}
\includegraphics[scale=0.4]{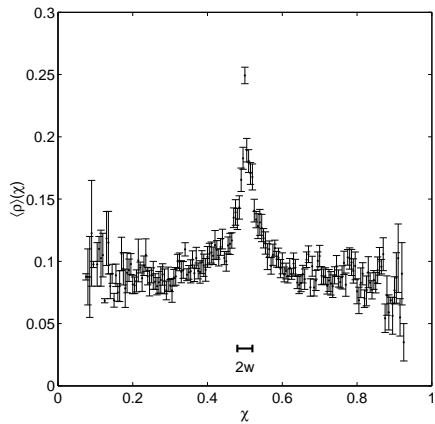}
\caption{\small Averaged damage density $\langle\rho\rangle(\chi)$. $L=100$, $R=1$.}
\label{fg:damage}
\end{center}
\end{figure}

This lack of the localization of damage is reflected in Fig.\ \ref{fg:damage}.
The damage density $\langle\rho\rangle(\chi)$ has been averaged in 
the $y$-direction as a function of the normalized coordinate transform 
$\chi=(x-x_c+L)/2L$, where the point $x_c$ is chosen as the one with the
maximum damage, and is located in practice at the final fracture line,
$x_c \simeq \bar{h}$.
After this shift, the average density is computed taking care that 
it is normalized correctly since the number of samples contributing for
each $\chi$ varies with the final crack location -  $x_c$ is a random 
variable. We also have added the average fracture line width $w(L=100)$ as a
comparison (from ref.~\cite{sep}). 
It can be seen that outside of the immediate vicinity of the
fracture process zone the damage is constant. Notice the error bars
of the data points, and that the data points located far away from the crack
line suffer from the presence of less data points as seen from the
error bars. It would be interesting to analyze in detail the functional
shape of $\langle\rho\rangle(\chi)$ in the proximity of the crackline,
$\chi = 0.5$. The implication of the results is that the density
can be written as a sum of a constant ($L$-dependent) background, 
and a term that has to decay (perhaps exponentially) within a finite
lengthscale from the crack. This decay length in turn may depend on $L$.

Such an observation is in contradiction
to the proposed ``self-consistent'' quadratic functional form, by
Hansen et al.\ \cite{Hansen}. This would imply 
$\langle\rho\rangle(\chi)=p_f-A\left(\frac{L(2\chi-1)}{l_x}\right)^2$, 
which is clearly not the case. In the light of the picture discussed
below about the internal dynamics of microcracks or avalanches, the
interpretation is that the final crack is formed here similarly to
weak disorder in a ``critical'' manner.  That is, once a damage
density sufficient for ``nucleation''\
is established the largest crack becomes unstable.
Prior to that the correlations
in the damage accumulated can for all purposes be neglected. This would
in turn to imply that the origin of the self-affine crack roughness
in fuse networks is {\em not dependent} on whether there is ``strong''
or ``weak'' disorder, as long as there are no infinitely strong
fuses, or as long as the process does not resemble e.g. percolation
due to the complete domination of zero-strength fuses.

\section{Avalanches}
Next we consider the correlations in the dynamics of individual
fuse failures. Recall that the MF-limit states that consecutive
ones should not be spatially correlated in any fashion; the 
opposite limit is given by the growth of a linear crack in 
which it is always the one adjacent to the crack tip to fail next. In the case
that the growing crack is ``rough'' one expects that the subsequent
failure takes place inside a fracture process zone, analogously 
to normal fracture mechanics, one of the follow-up questions
being how the size and the shape of this zone vary with system
size and as the crack grows \cite{rai}. The simplest quantities to compute,
to examine localization and spatial correlations between fractures,
are the 1d distances between consecutive fractures,
\begin{eqnarray}
\Delta x=|x_{i+1}-x_i|\\
\Delta y=|y_{i+1}-y_i|,
\end{eqnarray}
where $x_i$ and $y_i$ are the $x$- and $y$-coordinates of the $i$th fracture.
Another one choice is given by
the average distances between consecutive fractures belonging to the same 
avalanche (ie. induced by a single increment of the control parameter)
\begin{eqnarray}
\Delta x_{avalanche}=\frac{1}{\Delta-1}\sum_{i=1}^{\Delta-1}|x_{i+1}-x_i|\\
\Delta y_{avalanche}=\frac{1}{\Delta-1}\sum_{i=1}^{\Delta-1}|y_{i+1}-y_i|.
\end{eqnarray}
The MF theory predicts $P(\Delta x=k)=2(L-k)/L^2$ and $P(\Delta y=k)=2/L$,
if the boundary conditions used here are taken into account, and 
$\Delta$ is again the avalanche size measured in the number
of fuses broken during it.

\begin{figure}
\includegraphics[scale=0.35]{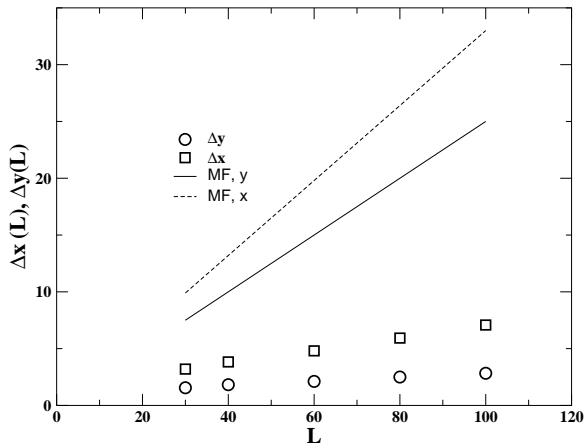}
\caption{\small Average one-dimensional distances $\langle\Delta x_{avalanche}\rangle$ and $\langle\Delta y_{avalanche}\rangle$ between consecutive broken fuses belonging to the same avalanche vs.\ linear dimension of system $L$. $R=1$. The solid lines are linear least squares fits with slopes $0.018$ for $\langle\Delta x_{avalanche}\rangle$ and $0.055$ for $\langle\Delta y_{avalanche}\rangle$. The dashed lines, linear with slopes $0.33$ for $\langle\Delta x_{avalanche}\rangle$ and $0.25$ for $\langle\Delta y_{avalanche}\rangle$, correspond to the distances predicted by mean-field theory. Statistical errors are smaller than or equal to the size of the data points. }
\label{fg:dvsL}
\end{figure}

Fig.\ \ref{fg:dvsL} depicts, as a comparison for the mean-field
results, the average distances in the $x$- and $y$-dimensions 
between consecutive broken fuses belonging to the same avalanche.
 $\langle\Delta x_{avalanche}\rangle$ and $\langle\Delta y_{avalanche}\rangle$, are shown, respectively, as a function of the system size. Both are linear
like in the MF theory, but with a smaller slope with $L$. This means,
that the damage created by a typical avalanche (microcrack creation,
crack advance etc.) {\em is localized} compared to the MF-prediction, 
but nevertheless the localization does not get stronger with $L$. 
One should note that the damage as such is almost volume-like.
The result is thus not surprising in the sense that a reduction
of the slope (sublinear behavior, say, $\langle\Delta x_{avalanche}\rangle
\sim L^\alpha$, with $\alpha < 1$)
would imply concomitant faster average crack growth,
which would be in contradiction with the damage scaling.

\begin{figure}
\includegraphics[scale=0.35]{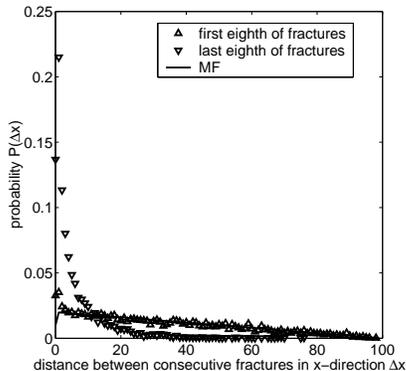}
\caption{\small Distribution of average distances in the $x$-dimension between consecutive broken fuses computed over the 1$^{\textrm{st}}$ and
8$^{\textrm{th}}$ 8$^{\textrm{th}}$ of the fractures of 20 realizations. 
$L=100$, $R=1$. The solid lines correspond to mean-field predictions.}
\label{Pddd}
\end{figure}

To understand in detail the dynamics of microcracks is a difficult 
task. This is since the growth dynamics is not local: the burned fuses
do {\em not} have to form connected clusters by any remotely easy
criterion. It is easy to comprehend that the driving force for
the localization is standard stress-enhancement, but as is true
for RFN's crack shielding and arrest (due to strong fuses,
in the early stages of fracture) play a role. One may set aside
for the sake of discussion the separation of the events into
avalanches, and just consider the distances between consecutive
burned fuses. Figure \ref{Pddd} demonstrates the difference 
between two probability distributions $P(\Delta x)$,
averaged over the first
1/8 of the typical failure process and the last, respectively,
for a fixed $L$. As one could expect, there is a peak in the
distribution (this holds for both $x$- and $y$-directions separately), 
which is greatly suppressed in the first part closest to the MF-limit.

Thus one may conclude, that there is a continuous cross-over
from purely MF-like behavior to a complicated non-local growth
dynamics. This is also exhibited by such distributions $P(\Delta x)$,
$P(\Delta y)$. The analysis of the detailed shape of the small-argument
part of the probability distributions would be an interesting
challenge. To first order, the result is a convolution of
a microcrack size distribution and the corresponding stress enhancement
factor, such that the distribution $P$ evolves according to the
growth indicated by the involved probability distributions. Given the
simple forms of say $P(\Delta x)$ for small arguments there might be
some hope for developing analytical arguments. 

When considered
as a function of $L$ it becomes immediately apparent why the
avalanche statistics resembles the MF-case so much. This is due
to two separate factors: first, the growth is clearly in the
sample case of Fig. \ref{Pddd} (or Fig. 1 again)
local over a certain lengthscale
(damage correlation length), $\xi_x$ or $\xi_y$. Second, one
should recall the scaling of the strength with  $L$: catastrophic
crack growth takes place earlier and earlier with respect to the
intensive variable, current. This means that the RFN's resemble,
in the thermodynamic limit, more and more the mean-field-case
in their fracture properties in spite of the stress- (or more
exactly current-) enhancements that the model contains.

\begin{figure}
\includegraphics[scale=0.35,angle=-90]{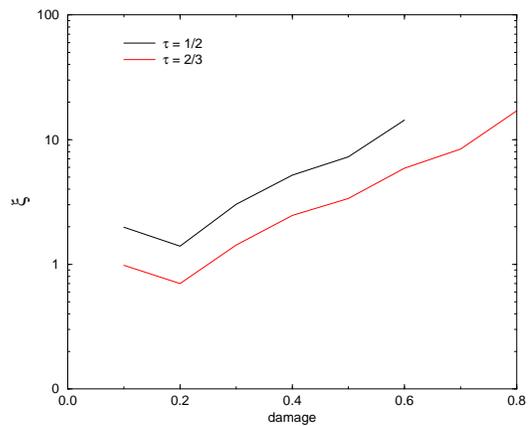}
\caption{\small The scaling of $\xi_y$ (see Eq.~(\ref{xiy})) with
increasing damage for $L=100$. The total number of broken bonds
has been divided into ten consecutive windows, and in each of these
$\langle d_y \rangle$ has been computed, and $\xi_y$ using Eq.~(\ref{xiy}).
For $\tau$ two guesses (1/2, 2/3) are used, note that $0<\tau<1$.}
\label{xi}
\end{figure}

Again, our data does not allow us to conclude firmly how such
correlation lengths behave for $\Delta x$ or $\Delta y$ small
- how the associated distributions $P$ would scale for small
arguments that is. One may however simply use an Ansatz
that $P \sim \Delta y^{-\tau}$ upto $\xi_y$, say and MF-like
for larger $\Delta y$ \cite{deplace}. This defines the correlation length
$\xi_y$ for a given damage density $\rho$. Using now the 
distribution $P$ allows to compute $\langle \Delta y \rangle$
and relate it to $\xi_y$, valid for such deviations from the
uncorrelated fracture process, that $\xi_y \ll \langle L/4 \rangle$.
The result is in analogy to Delaplace et al. \cite{deplace} that
\begin{equation}
\langle d_y \rangle = \frac{1-\tau}{4(2-\tau)}\frac{(2-\tau)L^2+8\tau\xi_y^2}
{(1-\tau)L+2\tau\xi_y}.
\label{xiy}
\end{equation}
In the opposite limit, $\xi_y \sim 1$ the correlations are
badly defined since the model is discrete. Figure \ref{xi}
shows an example of the ensuing scaling with different
guesses for $\tau$, for $L=100$. The main observation is
an exponential (perhaps) increase of $\xi_y$ with damage.
Again, note that with still larger system sizes the total
damage is diminished, which in turn implies that the maximal correlations
in the damage accumulation become weaker. Please observe
 that we have not
studied in detail the other possibility, $\xi_x$ since the main
interest lies with the correlations in the crack-growth
direction.

\section{Summary}

In this article we have studied the distributions and development
of damage in random fuse networks, with ``strong'' disorder. 
Our aim has been to understand possible deviations from mean-field
theory, and the associated correlations. This is of relevance both
as regards the statistical mechanics of fracture in general, and
in particular also the growth and formation of self-affine cracks.
In other words, we have concentrated on the ``approach to the critical
point'' if the failure transition is considered as an analogue of
ordinary phase transitions.

An analysis of the localization of damage both during the
fracture process and {\em a posteriori} reveals that in the
case studied the correlations are very weak, are formed
mostly in the last catastrophic phase of network failure,
after the maximum current $I_{max}$,
and do not have any global correlations. The localization
is centered in and around the final crack surface, or what may
be called as the total volume encompassed by a ``fracture
process zone''. We would like to note that this is in contrast
to the recent theory of self-organized damage percolation,
of Hansen {\em et al.} (\cite{Hansen2}) devised to explain the formation of
self-affine cracks in fuse networks, and in related experiments.
In particular it should be stressed that there is no evidence
of a global, non-trivial damage profile as contained in that
proposition. Recent numerical results of Nukala et al. \cite{kumar}
with much better statistics than what is the case
here or with other, earlier authors also seem to imply the same.

Since also in this particular case much of the
damage incorporated in the final crack is due to the ``last''
avalanche it seems then logical that the fracture surface
geometry is formed similarly to RFN's {\em with weak disorder},
for which there is no quasi-volume like damage, and one often
has just the propagation, and formation of the final crack.
Nevertheless of this complication,
the measured roughness exponents from such simulations
are close to that in the case here at hand, but note also that such
exponents are notoriously hard to measure numerically in finite
sized samples: more extensive work in this respect would certainly
be desirable.

The internal correlations of the avalanches become more and
more important as damage grows, but in line with the fact
that the statistics is close to the mean-field case 
the growth is never very far from the MF: for all phases studied
there are remote, broken fuses instead of ones localized close to
the last growth event. There is an associated lengthscale that
can be roughly defined based on the $x$- and $y$-dependent results,
but of course one could go further and look at the radial 
probability distribution $P(\vec{r})$, with 
$\vec{r} = (x_{i+1},y_{i+1}) - (x_i, y_i)$
(for which one would presumably need
still much larger systems to get decent averaging). One central
lesson is that localization will diminish with system size
due to the normal volume effect of strength, decreasing with $L$. 
In this respect,
fuse networks are not unique, and other simulation models of brittle
fracture should exhibit the same behavior. To conclude, even in our case with 
quite strong disorder the failure process consists of
weakly correlated damage growth and a final catastrophic
crack propagation phase, that induces a first-order drop
in the elastic modulus.

\acknowledgments
We are grateful to the Center of Excellence program
of the Academy of Finland for support.


\begin{thebibliography}{5}
\bibitem{Lockner}
D.A. Lockner {\em et al.}, 
	Nature {\bf 350}, 39 (1991).

\bibitem{Petri2}
A. Petri,
G. Paparo, A. Vespignani, A. Alippi, and M. Costantini,
	Phys. Rev. Lett {\bf 73}, 3423 (1994).

\bibitem{Ciliberto}
A. Guarino, A. Garcimartin, and S. Ciliberto,
Eur. Phys. J. B {\bf 6}, 13 (1998);
A. Garcimartin {\em et al.}, 
Phys. Rev. Lett {\bf 79}, 3202 (1997).

\bibitem{penn}
L.C. Krysac and J.D. Maynard, 
	Phys. Rev. Lett. {\bf 81}, 4428 (1998).

\bibitem{sal1}
L.I. Salminen, A.I. Tolvanen, and M.J. Alava,
	Phys. Rev. Lett {\bf 89}, 185503  (2002).

\bibitem{Mandel} B.~B.~Mandelbrot, D.~E.~Passoja, and A.~J.~Paullay,
 Nature (London) {\bf 308}, 721 (1984).

\bibitem{Bourev} E.\ Bouchaud, J.\ Phys.\ Cond.\ Mat.\ {\bf 9}, 4319 
(1997).

\bibitem{bouprl} P.\ Daguier, B.\ Nghiem, E.\ Bouchaud, and F.\ Creuzet,
Phys.\ Rev.\ Lett.\ {\bf 78}, 1062 (1997).

\bibitem{hansen}
M. Kloster, A. Hansen, and P.C. Hemmer, 
	Phys. Rev. {\bf E56}, 2615 (1997).

\bibitem{Zapperi} S.~Zapperi {\em et al.},
Phys. Rev. Lett. {\bf 78}, 1408 (1997). 
\bibitem{review} Chapters 4-7 in {\it Statistical models for the
fracture of disordered media}, ed. H.\ J.\ Herrmann and S.\ Roux,
(North-Holland, Amsterdam, 1990).

\bibitem{Dux} P.\ M.\ Duxbury, P.\ L.\ Leath, and P.\ D.\ Beale, 
Phys.\ Rev.\ B {\bf 36}, 367 (1987); P.\ M.\ Duxbury, P.\ L.\ Leath, and 
P.\ D.\ Beale, Phys.\ Rev.\ Lett.\ {\bf 57}, 1052 (1986).

\bibitem{Kahng}
B.\ Kahng, G.\ G.\ Batrouni, S.\ Redner, L.\ de Arcangelis, and
H.\ J.\ Herrmann, Phys.\ Rev.\ B \textbf{37}, 7625 (1988).

\bibitem{Hansen} A.\ Hansen, E.\ L.\ Hinrichsen, and S.\ Roux,
Phys.\ Rev.\ Lett.\ {\bf 66}, 2476 (1991).

\bibitem{rai}
V.\ R\"ais\"anen, M.\ Alava, E.\ Sepp\"al\"a, and
 P.\ M.\ Duxbury, Phys.\ Rev.\ Lett.\ \textbf{80}, 329 (1998).

\bibitem{sep}
E.\ T.\ Sepp\"al\"a, V.\ I.\ R\"ais\"anen, and
M.\ J.\ Alava, Phys.\ Rev.\ E, \textbf{61}, 6312 (2000).

\bibitem{Kertesz} J.\ Kert{\'e}sz, V.\ K.\ Horvath, and F.\ Weber, Fractals
{\bf 1}, 67 (1993).

\bibitem{Maloy} T. Engoy, K.J. Maloy, and A. Hansen,
Phys. Rev. Lett. {\bf 73}, 834 (1994).

\bibitem{Rosti}
J. Rosti {\em et al.},
Eur. Phys. J. {\bf B19}, 259 (2001).


\bibitem{sal2}
L.I. Salminen, M.J. Alava, and K.J. Niskanen,
Eur. Phys. J. {\bf B32}, 369 (2003).

\bibitem{Zapperi2}
S.\ Zapperi, P.\ Ray, H.\ E.\ Stanley, and
A.\ Vespignani, Phys.\ Rev. E, \textbf{59}, 5049 (1999).

\bibitem{pitka} 
For 3d RFN's the situation is not
so clear-cut: V.\ I.\ R\"ais\"anen, M.\ J.\ Alava, and
R.\ M.\ Nieminen, Phys.\ Rev.\ B {\bf 58}, 14288 (1998).

\bibitem{minozzi}
M. Minozzi, G.   Caldarelli, L. Pietronero, and
S. Zapperi, Eur. Phys. J. {\bf B36}, 203 (2003).

\bibitem{Hansen2}
A.\ Hansen and
 J.\ Schmittbuhl, Phys.\ Rev.\ Lett.\ \textbf{90}, 045504 (2003).

\bibitem{ramstad}
T. Ramstad, J.O.H. Bakke, J. Bjelland, T. Stranden, and
A. Hansen,
cond-mat/0311606.

\bibitem{dipolar}
M. Barthelemy, R. da Silveira, and H. Orland,
Europhys. Lett. {\bf 57}, 831 (2002).

\bibitem{Curtin}  W.A. Curtin,
	Phys. Rev. Lett. {\bf 80}, 1445 (1998).

\bibitem{van}
P. Van, C. Papenfuss, and W. Muschik,
	Phys. Rev. {\bf E62}, 6206 (2000).
\bibitem{deplace}
A. Delaplace, G. Pijaudier-Cabout, and S. Roux,
J. Mech. Phys. Solids {\bf 44}, 99 (1996)


\bibitem{kumar}
P. K. V. V. Nukala, S. Smiunovic, and S. Zapperi,
cond-mat/0311284.

\end{thebibliography}
\end{document}